# Possible structural origin of superconductivity in Sr-doped Bi₂Se₃

## Authors:

Zhuojun Li[1,6]*, Meng Wang[1,6,8]*, Dejiong Zhang[2]*, Nan Feng[3]*, Wenxiang Jiang[4], Chaoqun Han[4], Weijiong Chen[4], Mao Ye[1,6], Chunlei Gao[5,7], Jinfeng Jia[4,7], Jixue Li[2], Shan Qiao[1,6,8], Dong Qian[4,7], Ben Xu[3†], He Tian[2†], Bo Gao[1,6,8†],

## Affiliations:

[1]CAS Center for Excellence in Superconducting Electronics (CENSE), Shanghai 200050, China

[2]Center of Electron Microscopy and State Key Laboratory of Silicon Materials, School of Materials Science and Engineering, Zhejiang University, Hangzhou 310027, China

[3]State Key Laboratory of New Ceramics and Fine Processing, Department of Materials Science and Engineering, Tsinghua University, Beijing, 100084, P. R. China

[4]Key Laboratory of Artificial Structures and Quantum Control (Ministry of Education), School of Physics and Astronomy, Shanghai Jiao Tong University, Shanghai 200240, China

[5]State Key Laboratory of Surface Physics and Department of Physics, Fudan University, Shanghai 200433, China

[6]State Key Laboratory of Functional Materials for Informatics, Shanghai Institute of Microsystem and Information Technology, Chinese Academy of Sciences, 865 Changning Road, Shanghai 200050, China.

[7]Collaborative Innovation Center of Advanced Microstructures, Nanjing 210093, China

[8]University of Chinese Academy of Sciences, Beijing 100049, China

*These authors contribute equally to this work.

†Corresponding authors. E-mail: xuben@mail.tsinghua.edu.cn, hetian@zju.edu.cn, bo_f_gao@mail.sim.ac.cn

# Abstract:

Doping bismuth selenide ($Bi_2Se_3$) with elements such as copper and strontium (Sr) can induce superconductivity, making the doped materials interesting candidates to explore potential topological superconducting behaviors. It was thought that the superconductivity of doped $Bi_2Se_3$ was induced by dopant atoms intercalated in van der Waals gaps. However, several experiments have shown that the intercalation of dopant atoms may not necessarily make doped $Bi_2Se_3$ superconducting. Thus, the structural origin of superconductivity in doped $Bi_2Se_3$ remains an open question. Herein, we combined material synthesis and characterization, high-resolution transmission electron microscopy, and first-principles calculations to study the doping structure of Sr-doped $Bi_2Se_3$. We found that the emergence of superconductivity is strongly related with n-type dopant atoms. Atomic-level energy-dispersive X-ray mapping revealed various n-type Sr dopants that occupy intercalated and interstitial positions. First-principles calculations showed that the formation energy of a specific interstitial Sr doping position depends strongly on Sr doping level. This site changes from a metastable position at low Sr doping level to a stable position at high Sr doping level. The calculation results explain why quenching is necessary to obtain superconducting samples when the Sr doping level is low and also why slow furnace cooling can yield superconducting samples when the Sr doping level is high. Our findings suggest that Sr atoms doped at interstitial locations, instead of those intercalated in van der Waals gaps, are most likely to be responsible for the emergence of superconductivity in Sr-doped $Bi_2Se_3$.

# I. INTRODUCTION

Majorana fermions, named after Italian theoretical physicist Ettore Majorana, are hypothetical particles that are their own antiparticles. Recently, Majorana fermions have sparked substantial research interest in condensed matter physics because, after more than eighty years of searching in the realm of high energy physics [1], it turns out that these exotic particles may exist in topological superconductors (TSCs), in which a bulk superconducting gap coexists with gapless surface states [2,3]. The quasiparticle excitations of these surface states are expected to be Majorana fermions, whose discovery, if confirmed, will be a great triumph of quantum theory. In addition, because these surface states are topologically protected and thus immune to local perturbations, they are considered to be ideal building blocks for future fault-tolerant quantum computing [4,5]. There have been many experimental attempts to realize TSCs. Possible signatures of Majorana quasiparticles like a zero-bias conductance peak (ZBCP) have been observed in various systems such as a semiconducting nanowire with strong spin–orbit coupling proximately coupled to a superconducting electrode [6,7], ferromagnetic atomic chains placed on an s-wave superconductor [8,9], a topological insulator (TI) thin film coupled with an s-wave superconducting film [10-12], and superconducting doped TIs [13] and topological crystalline insulators (TCIs) [14].

Doping TIs and TCIs to induce superconductivity has been considered a promising way to create TSCs. This approach has the advantage of avoiding the difficulty of fabricating a transparent superconductor–semiconductor interface. In addition, because the surface states of TSCs are robust against non-magnetic impurities, the non-magnetic dopant atoms will not harm the induced topological superconductivity. To date, doping-induced superconductivity has been reported for Cu-, Sr- and Nb-doped $Bi_2Se_3$ [15-17], Pd- and Tl- doped $Bi_2Te_3$ [18,19], and In-doped SnTe [20]. Extensive effort has been devoted to searching for signatures of Majorana fermions in these doped materials [13,21-37], but experimental results are not consistent. For example, Cu-doped $Bi_2Se_3$ is the most extensively studied superconducting doped TI. Point contact spectroscopy of this material showed a ZBCP, suggesting unconventional superconductivity [13,24]; direct-current (DC) magnetization [25] and upper critical field measurements [26] also pointed to spin–triplet superconductivity. However, ultra-low temperature scanning tunneling spectroscopy of Cu-doped $Bi_2Se_3$ did not reproduce the ZBCP but revealed a fully gapped conventional s-wave superconducting structure without mid-gap states [27].

To obtain smoking-gun type evidence for Majorana fermions in superconducting doped TIs/TCIs, transport measurements involving reliable tunneling junctions are strongly desired, which requires fabricating doped samples in thin-film or nanostructured form. Unfortunately, such tasks are extremely difficult. Attempts to grow superconducting doped TIs or TCIs in thin-film form have not yet been successful. The widely used mechanical exfoliation technique hardly yielded superconducting nanoflakes from bulk superconducting samples. There are only few reports about superconducting doped TIs/TCIs nanoflakes grown by chemical vapor deposition or complex patterning/annealing process [38-40]. The difficulty of fabricating low-dimensional superconducting doped TIs thus raises the question of the structural origin of superconductivity in doped TIs. For Cu-doped $Bi_2Se_3$, it was assumed that dopants inside van der Waals (VDW) gaps are responsible for the emergence of superconductivity because considerable c-axis lattice expansion was observed in superconducting samples [15]. However, various experiments have shown that intercalation of dopants inside VDW gaps may not necessarily make Cu-doped $Bi_2Se_3$ superconducting [13,41]. Attempts to fabricate superconducting Cu-doped $Bi_2Se_3$ thin films have not yet been successful, although Cu intercalation in VDW gaps can be easily realized through molecular beam epitaxy [41] or electrochemical intercalation [13]. Similar controversy also exists for Sr-doped $Bi_2Se_3$. Although a recent scanning tunneling spectroscopy study identified superconducting regions, it was difficult to tell whether the doped Sr atoms were intercalated in VDW gaps or inserted inside quintuple layers [30].

Consequently, the lack of information about the structural origin of superconductivity in doped TIs currently impedes the efforts to obtain high quality superconducting doped TIs in thin-film or nanostructured form, and hinders the understanding of the superconducting mechanism in these materials. Therefore, there is an urgent need to clarify which doping location is responsible for the emergence of superconductivity in doped TIs. Here, we chose to study Sr-doped $Bi_2Se_3$ in an attempt to determine the doping location that induces superconductivity because of the ease of fabricating superconducting samples with high superconducting volume fraction.

## II. PROCEDURES

Sr-doped $Bi_2Se_3$ single-crystal samples were synthesized by melting mixtures of high-purity bismuth, selenium, and strontium with nominal atomic ratios of 2:3:$x$, where the nominal Sr concentration $x$ was between 0.03 and 0.25. Each mixture was prepared in a nitrogen-filled glove box and then sealed in a quartz ampoule. The ampoule was kept at 850 °C for 24 h, and then cooled to 620 °C at a rate of 3 °C/h. Most samples were then quenched in ice water. A few non-quenched samples were synthesized by cooling the ampoules from 620 °C to room temperature in the furnace.

Powder X-ray diffraction (XRD) patterns were obtained using a powder X-ray diffractometer (DX-2700, Dandong Haoyuan Instrument Co., Ltd. China) at 40 kV and 40 mA over the 2θ range of 10°–70°. Electrical measurements were performed in a cryostat (helium-4) and dilution fridge using the DC and lock-in technique. Diamagnetic measurements were carried out in a commercial Quantum-Design MPMS system. Cross-sectional samples for high-resolution transmission electron microscopy (HRTEM) were prepared using a dual-beam microscope (FIB, Quanta 3D, FEG, FEI) with Ga-ion milling and precision ion-polishing system (Gatan 691) with Ar-ion milling. The structural defects in samples were examined with a transmission electron microscope (FEI TITAN Cs-corrected ChemiSTEM) operated at an acceleration voltage of 200 kV to avoid knock-on damage. High-angle annular dark-field scanning transmission electron microscopy (HAADF-STEM) was conducted using a spherical aberration probe corrector to achieve a spatial resolution of up to 0.08 nm. The ChemiSTEM energy-dispersive X-ray (EDX) spectrometer provided outstanding sensitivity to determine elements with atomic resolution.

## III. RESULTS AND DISCUSSION

### A. Evolution of superconducting properties with Sr nominal concentration

We investigated the evolution of superconducting properties of bulk samples of Sr-doped $Bi_2Se_3$ with various Sr nominal concentrations. Figure 1(a) depicts the temperature-dependent resistivity of a typical superconducting sample. The sample exhibits a superconducting transition at around 2.8 K. Figure 1(b) shows the field-cooling and zero-field-cooling (ZFC) diamagnetic measurement results obtained for the same sample. The superconducting volume fraction of the sample reaches nearly 70% at 1.8 K, which is much higher than the reported value for Cu-doped $Bi_2Se_3$ [42]. The dependence of the superconducting volume fraction on electron density is plotted in Fig. 2(a). The superconducting volume fraction was deduced from the ZFC diamagnetic curve measured at 1.8 K and the electron density was obtained from Hall measurements. The left panel of Fig. 2(a) reveals that the samples with low electron density have a vanishing superconducting volume fraction. As the electron density increases, a positive correlation between the superconducting volume fraction and electron density is observed, as seen in the right panel of Fig. 2(a). Because the electron density is strongly affected by Sr doping, the positive correlation between these parameters suggests that n-type Sr dopants influence the emergence of superconductivity in Sr-doped $Bi_2Se_3$.

The dependence of electron density on the nominal Sr concentration of the sample is presented in Fig. 2(b). The left panel shows that low nominal Sr concentration generally leads to low electron density regardless of the cooling method used to synthesize the sample. The independence of electron density and cooling method is also found for the samples with high Sr nominal concentration (right panel). An exception is observed in the region near an Sr nominal concentration of 0.05, where the quenching technique yields samples with high electron density and furnace cooling yields samples with low electron density. In this critical Sr concentration region, samples synthesized using different cooling methods show distinct superconducting behaviors. As illustrated in Fig. 2(c) and (d), the quenched samples can reach zero resistance and possess large ZFC superconducting volume fractions. By contrast, the furnace-cooled samples do not reach zero resistance and show weak diamagnetic behavior. The important role played by quenching in obtaining superconducting samples suggests that the superconductivity of Sr-doped $Bi_2Se_3$ should originate from an energetically metastable Sr doping location.

## B. High-resolution transmission electron microscopy investigation

As mentioned above, the positive correlation between the superconducting volume fraction and electron density of the samples implies that the superconductivity originates from n-type Sr doping. It is known that the Bi substitution defect $Sr_{Bi}^{-1}$ is a p-type donor [43], which therefore cannot induce superconductivity. A recent scanning tunneling spectroscopy study suggested that the superconductivity originates either from Sr dopant atoms intercalated in VDW gaps or inserted into quintuple layers [30]. To obtain a further information about the Sr dopant atoms and their local environments, we used HAADF-STEM combined with EDX mapping to inspect the Sr-doped $Bi_2Se_3$ samples. We chose $Sr_{0.05}Bi_2Se_3$ samples because higher Sr nominal concentration yielded an impurity phase (as indicated by powder XRD analysis, which is provided in the Supplementary Materials). Figure 3(a) shows a HAADF-STEM image of a bulk $Sr_{0.05}Bi_2Se_3$ sample. The image clearly reveals the typical structure of quintuple layers separated by VDW gaps. Because the intensity in HAADF-STEM images is roughly proportional to $Z^2$ ($Z$ is the atomic number), the columns of Bi atoms are of higher intensity than those of Se atoms. We can barely discern dopant atoms intercalated in VDW gaps or at any other location because of the low Sr concentration of the sample and low atomic number of Sr.

To locate the Sr dopant atoms, we used EDX mapping to perform an atom-by-atom elemental analysis of the sample. Figure 3(b) and (c) display the EDX maps of the Bi/Se mixture in pristine bulk $Bi_2Se_3$ and the ideal structure model of $Bi_2Se_3$, respectively. The EDX maps agree well with the model. Compared to HAADF-STEM imaging, EDX mapping has the advantage of being able to distinguish very weak signals originating from doped Sr. Figure 3(d) shows the EDX maps obtained for the same Sr-doped $Bi_2Se_3$ bulk sample and Fig. 3(e) is a schematic of Sr-doped $Bi_2Se_3$ corresponding to the experimental structure in Fig. 3(d). These maps reveal that the Sr dopant atoms can occupy the following locations in the $Bi_2Se_3$ lattice: i) The Sr dopant atoms (indicated by white arrows) align well with the blue Bi atoms in the same column, forming the Bi substitution defect $Sr_{Bi}^{-1}$; ii) Other Sr dopant atoms (indicated by solid circles) occupy the *inner* and *outer* interstitial sites between the Bi and Se layers, forming interstitial defect $Sr_i^{+1}$; iii) The remaining Sr dopant atoms (indicated by black arrows) reside near an Se atom, probably forming another type of $Sr_i^{+1}$ interstitial defect in the Se layer (the substitution of Se by Sr is energetically less favorable than that of Bi substitution). It is noteworthy that even in the EDX maps, it was still difficult to observe the trace of Sr dopant atoms between neighboring columns of Se (inside VDW gaps). It does not mean that Sr dopant atoms intercalated in VDW gaps are absent. As shown in the red oval in Fig. S3 in the Supplementary Materials, Sr dopant atoms intercalated in VDW gaps do exist. However, the intensity of EDX mapping is proportional to the number of atoms superposed along the direction of observation. Because of the large layer spacing between adjacent quintuple layers, metal dopant atoms can travel freely within VDW gaps [44,45]. The electron beam illumination may make the intercalated Sr dopant atoms even more mobile. Therefore, at any given position inside VDW gaps, the number of the intercalated Sr atoms superposed along the direction of observation is limited, making it difficult to discern these dopant atoms in the EDX mapping.

## C. First-principles calculations

HRTEM observations revealed various Sr doping locations inside the $Bi_2Se_3$ lattice. We then used first-principles calculations to determine the formation energy $E_{form}$ corresponding to each Sr doping location. $E_{form}$ is defined as

$$E_{form} = (E_{tot}(TM) - E_{tot}(bulk) - \sum_i n_i \mu_i)/N, \tag{1}$$

where $E_{tot}(TM)$ is the total energy of the doped system, $E_{tot}(bulk)$ is the total energy of the undoped system, $n_i$ is the number of atoms added to or removed from the system, $\mu_i$ is the corresponding chemical potential of the added/removed atoms, and $N$ is the total number of atoms in the system. We first considered the situation with only one Sr atom in a 2×3×1 (multiples of the unit cell along the a-b-c directions) Bi₂Se₃ supercell. Table I lists $E_{form}$ of four doping locations: inside the VDW gap, Bi substitution, between Bi and inner Se layers (denoted Se2-Bi), and Se substitution. The calculation results for other doping structures are given in the Supplementary Materials. Figure 4 presents schematics of the pristine and four Sr-doped Bi₂Se₃ structures. Our calculations show that most of the interstitial Sr dopant atoms will move into VDW gaps after relaxation. Se2-Bi interstitial doping is an exception because Sr dopant atoms remain inside the quintuple layers after relaxation. Considering $E_{form}$, Se substitution is very unlikely to occur. In addition, Se2-Bi interstitial doping is energetically less stable than Bi substitution doping and VDW intercalation doping. Interestingly, if two Sr atoms are added into a supercell of 2×3×1, $E_{form}$ of Se2-Bi interstitial doping becomes comparable with those of Bi substitution doping and VDW intercalation doping, meaning this position changes from a metastable position to a more stable one with increasing Sr content.

TABLE I. Formation energies for various Sr doping locations in Bi₂Se₃

| Doping location | VDW | Bi subs | Se2-Bi | Se subs |
|---|---|---|---|---|
| E¹_form (meV) | − 22.292 | − 22.160 | − 13.723 | 24.581 |
| E²_form (meV) | − 49.15 | − 40.90 | − 42.34 | N/A |

E¹_form and E²_form correspond to one and two Sr dopant atoms in a 2×3×1 supercell,

respectively.

We now discuss which doping location is most likely responsible for the emergence of superconductivity in the Sr-doped $Bi_2Se_3$ samples. First, the positive correlation between the superconducting volume fraction and electron density of the samples suggests the emergence of superconductivity in Sr-doped $Bi_2Se_3$ is induced by n-type dopants. Therefore, Sr dopant atoms that substitute Bi atoms are not likely to be the origin of superconductivity. The remaining candidates are Sr dopants intercalated in the VDW gaps and those at the interstitial sites. A previous study also indicated that the structural origin of superconductivity was these locations [30]. Second, we found that for samples with low Sr nominal concentration, quenching is necessary to produce superconducting samples. This is similar to the role played by quenching in obtaining superconducting Cu-doped $Bi_2Se_3$ samples. The need for quenching suggests that the superconductivity is caused by a metastable Sr doping location. However, we also found that for samples with high Sr concentration, quenching is no longer required to obtain superconducting samples; slow furnace cooling could also yield superconducting samples.

Our first-principles calculations showed that Sr intercalation inside VDW gaps is always a stable doping location. Therefore, this site can hardly justify the importance of quenching to obtain superconducting Sr-doped $Bi_2Se_3$ samples. By contrast, if we consider only one Sr dopant atom in a 2×3×1 supercell, the Se2-Bi interstitial location has higher $E_{form}$, thus is less stable and is more likely to be a candidate for the structural origin of superconductivity than Sr intercalation inside VDW gaps. More importantly, when two Sr dopant atoms are placed in the same 2×3×1 supercell, Se2-Bi doping changes from an energetically less stable location to a more stable one. This result qualitatively explains why quenching is not a necessary step to provide superconducting samples when the Sr nominal concentration is high. Considering all these findings, we think that the Se2-Bi interstitial doping between Bi and the inner Se layer is more likely to be the origin of superconductivity in Sr-doped $Bi_2Se_3$ than Sr intercalation inside VDW gaps.

## IV. SUMMARY

We used materials synthesis and characterization, atom-resolved EDX mapping, and first-principles calculations to investigate the structural origin of superconductivity in Sr-doped $Bi_2Se_3$. We found that instead of the usually assumed Sr dopant atoms intercalated in VDW gaps, it was Sr atoms inserted between Bi and inner Se layers that should be most likely responsible for the emergence of superconductivity. The first-principles calculations indicated that this interstitial Sr doping location is energetically metastable when the Sr doping level is low, which explains the importance of quenching to obtain superconducting samples. Our findings may help to identify the origin of superconductivity in other doped TIs and provide guidance to grow superconducting doped TI thin films, which have potential applications in the detection of Majorana fermions.

## Acknowledgements

We thank D.W Shen and W. Li for helpful discussions. This work was supported by by National Natural Science Foundation of China under Grant No. 11374321, No.

11474249, No. 51672155, No. 51202232, No. U1632272, No. U1632266, No. 11574201, No. 11521404, No. 11227902, No. 11674063 and No.11374206; by Ministry of Science and Technology of China under Grant No. 2016YFA0301003; by the National 973 Program of China under Grant No.2015CB654901 and No. 2013CB632506, by The National Key Research and Development Program of China under Grant 2017YFB0703100, No. 2016YFA0201003 and No. 2016YFA0300904. H.T. acknowledges support from Young 1000 Talents Program of China. D.Q. acknowledges support from the Changjiang Scholars Program.

**Fig.1**

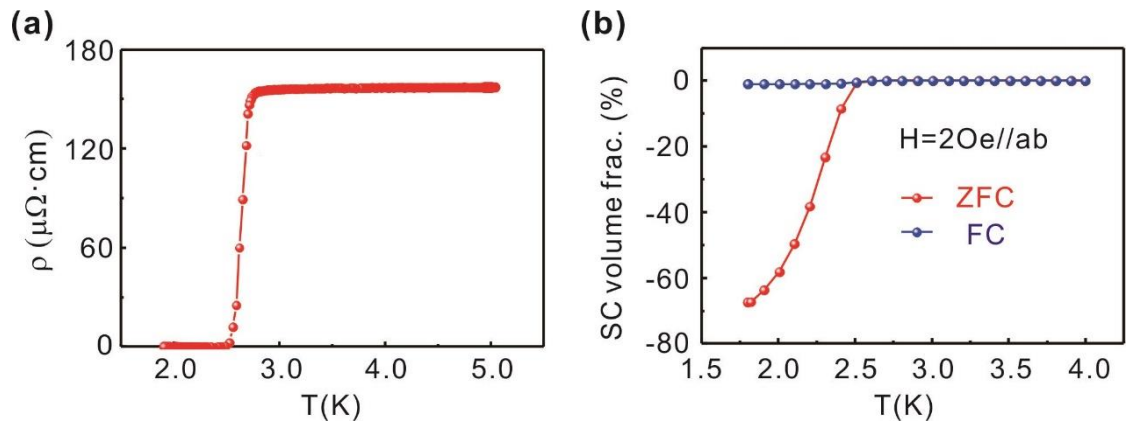

FIG. 1. Sample characterization. (a) Temperature dependence of the resistivity of a bulk $Sr_{0.05}Bi_2Se_3$ sample. (b) Diamagnetic measurements for the same bulk sample.

**Fig.2**

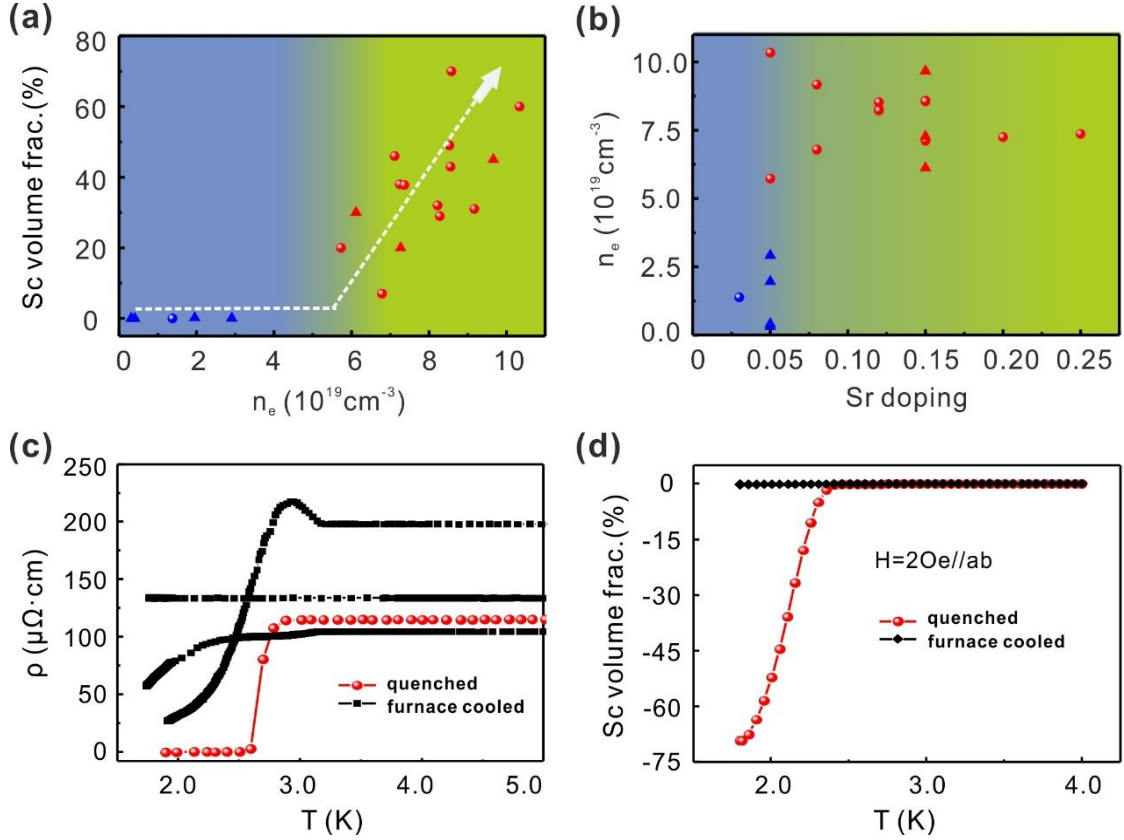

FIG. 2. Electronic properties of bulk Sr-doped Bi₂Se₃ samples as a function of nominal Sr concentration. Circles and triangles show the results for quenched and furnace-cooled samples, respectively. Superconducting samples are shown in red and non-superconducting samples in blue. (a) Superconducting volume fraction as a function of electron density. Left panel (blue background): samples with low electron density; right panel (yellow background): samples with high electron density. (b) Electron density as a function of nominal Sr concentration. Left panel (blue background): low nominal Sr concentration; right panel (yellow background): high Sr nominal concentration. (c) and (d) Electrical and diamagnetic characteristics of quenched and furnace-cooled Sr-doped Bi₂Se₃ bulk samples. Diamagnetic measurements were performed with the field oriented parallel to the *ab* plane under zero-field-cooling conditions.

**Fig.3**

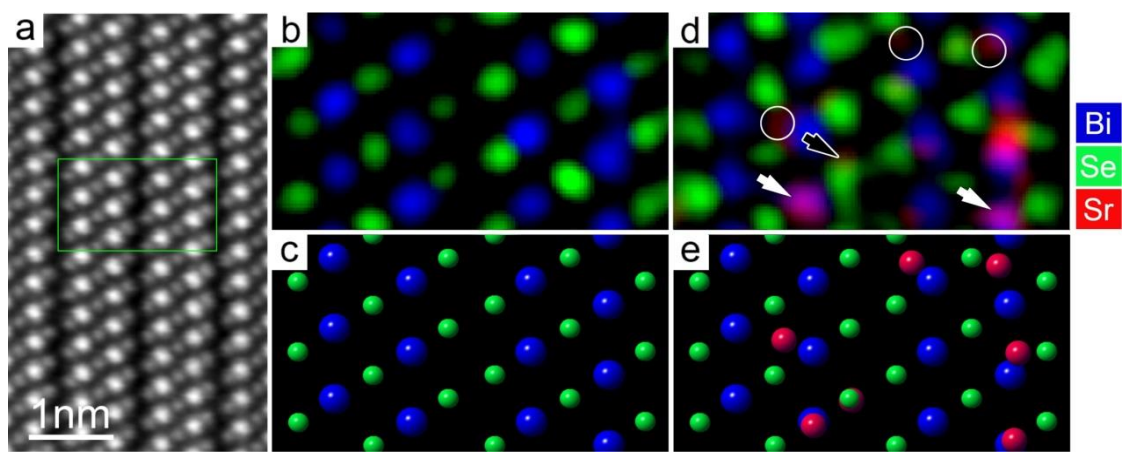

FIG. 3. Structural and chemical characterization of Sr-doped $Bi_2Se_3$ samples by STEM. (a) HAADF-STEM images of quenched Sr-doped $Bi_2Se_3$ with a nominal Sr doping level of $x$=0.05. (b) Atomic-resolution EDX element map of a pristine $Bi_2Se_3$ bulk sample. (c) Schematic showing the atomic structure of $Bi_2Se_3$. (d) EDX map of an $Sr_{0.05}Bi_2Se_3$ bulk sample. (e) Schematic of an Sr-doped $Bi_2Se_3$ sample corresponding to the experimental structure in (d). Bi, Se, and Sr atoms are shown in blue, green, and red, respectively.

**Fig. 4**

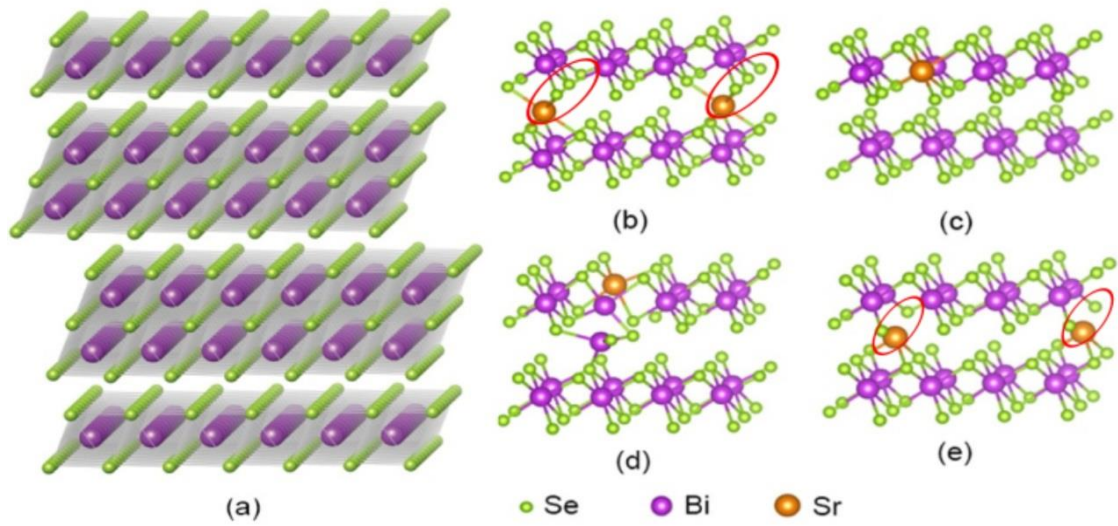

(a)

(b)

(c)

(d)

(e)

○ Se    ○ Bi    ○ Sr

FIG. 4. Structure of Bi$_2$Se$_3$. Schematics of (a) pristine Bi$_2$Se$_3$ and Sr dopant atoms in the Bi$_2$Se$_3$ lattice at (b) VDW, (c) Bi sub, (d) Se2-Bi, and (e) Se sub positions, respectively. The difference between (b) and (e) is highlighted by red ovals in the figures.